# Microstructured Electrode-Piezopolymer Interface for Ultrasound Transducers with Enhanced Flexibility and Acoustic Performance

Spencer Hagen[§], Dulcce A Valenzuela[§], Parag V Chitnis*, Shirin Movaghgharnezhad*

Department of Bioengineering, George Mason University, Fairfax, VA 22030, USA

[§]These authors contributed equally to this work.

* Corresponding authors.
*E-mail addresses:* Shirin Movaghgharnezhad (smovaghg@gmu.edu), Parag V Chitnis (pchitnis@gmu.edu)

Ultrasound transducers made from rigid piezoceramics are difficult to adapt for wearable or conformal applications. Piezopolymer-based transducers offer a practical alternative; however, most existing studies focus on piezoelectric materials, while the influence of electrode material and electrode–polymer interface remains underexplored. This study leverages different interface-engineering strategies to examine the influence of electrode–piezopolymer interface morphology on piezoelectric, dielectric, and acoustic behavior in flexible transducers. Devices were fabricated using silver (Ag), gold (Au), graphene flakes (GF), laser-induced graphene (LIG), and Au-decorated LIG electrodes, enabling comparison across interfacial architectures. LIG-based transducers showed strong acoustic and piezoelectric output due to partial infiltration of the piezopolymer into the porous LIG network, which enhances interfacial contact and stress transfer. Au-based transducers achieved comparable acoustic output. In contrast, dense Ag electrodes and layered GF films provided limited coupling, resulting in reduced electromechanical response. LIG-based transducers exhibited the highest flexibility and durability, retaining stable performance after 10,000 bending cycles and an eight-week aging study, whereas GF, Ag, and Au devices degraded under bending, and Ag electrodes declined over time. These findings demonstrate that engineering the electrode–polymer interface is critical for high-performance flexible ultrasound transducers and identify LIG as a strong candidate for wearable imaging applications.

## 1. Introduction

Ultrasound is widely used for medical diagnostics and industrial inspection because it is noninvasive, accurate, and capable of probing deep into structures. [1–3] Conventional ultrasound transducers (UST), typically fabricated from rigid piezoceramics, are limited by their brittleness, bulk, and poor conformity to nonplanar surfaces. These constraints hinder accurate imaging of curved anatomical regions or confined structures such as pipelines, while also restricting use in dynamic clinical settings. The rigid contact surfaces of these probes often leave air gaps when applied to irregular geometries, reflecting a substantial fraction of acoustic energy and causing signal distortion or loss. [4] Although coupling gels or water are commonly used to reduce this problem, they can introduce additional energy losses due to impedance mismatch and may obscure weaker echoes when overapplied.[5] The fixed foorprint and the mechanical rigidity of these probes further limit access to tight spaces and complicate continuous imaging of moving tissues. These limitations have driven increasing interest in flexible and wearable ultrasound systems that employ compliant substrates and thin-film piezopolymers to improve acoustic coupling, conformability, and performance for a range of practical applications.

Flexible ultrasound transducers (FUSTs) have emerged over the past decade as a promising approach to overcoming the limitations of rigid probes. [6–14] In a typical piezoelectric UST, a piezoelectric layer is sandwiched between top and bottom metal-based electrodes, enabling the bidirectional conversion of electrical signals and acoustic waves. Building on this structure, most FUSTs rely on compliant substrates to introduce mechanical flexibility without compromising device performance. The choice of substrate is particularly important, as it determines both the device's mechanical compliance and its long-term integrity. Elastomers such as polydimethylsiloxane (PDMS) and Ecoflex are widely used because of their stretchability and low modulus, which allow conformal contact with skin or other curved surfaces and stable operation during deformation. [1,15–17] In addition to elastomers, substrates such as polyurethane and thermoplastic polymers, including polyimide (PI) [12,18], polyethylene terephthalate (PET), and polyethylene naphthalate (PEN), have been widely adopted owing to their mechanical robustness and biocompatibility. [11,14,19,20] The use of these substrates is typically accompanied by flexible piezopolymers, most notably polyvinylidene fluoride (PVDF) and its composites, [16,17,21] which have been extensively investigated for FUSTs because of their inherent flexibility, low acoustic impedance, and broad bandwidth. While flexible substrates and piezoelectric layers have been widely studied, electrode design remains comparatively underexplored, despite its critical role in maintaining both electrical performance and mechanical compliance.

To preserve device flexibility while maintaining reliable electrical functionality, thin metal films or conductive inks [12] are commonly used as bottom electrodes. Gold (Au) [22], Silver (Ag) [12], and copper (Cu) [23] are often chosen due to their high conductivity and compatibility with microfabrication processes. Several design strategies have been explored to impart flexibility into metal electrodes. For example, copper films patterned into serpentine geometries on PI substrates can retain electrical continuity under bending. [23] Molybdenum (Mo) electrodes integrated with scandium-doped aluminum nitride (AlN) thin films have been shown to provide both electrical performance and mechanical stability. [24] In addition, flexible metal foils, such as stainless steel (SS), can serve as both substrates and electrodes, simplifying device structure. [25] While these methods enable functional, flexible devices, they often require complex fabrication steps and can suffer from a mechanical mismatch between stiff metals and compliant polymers. [26,27] Such a mismatch leads to reduced compliance and eventual delamination during bending, a common issue in flexible electronics. [28] . This metal-polymer mismatch is a recognized challenge in flexible electronics, motivating the development of alternative electrode materials and architectures that match soft substrates while maintaining high electrical conductivity. [14,29]

One solution has been the use of intrinsically compliant or porous conductors, which mitigate interfacial stresses by distributing strain more uniformly across the interface. Metallic nanowires (NWs), for example, combine high conductivity with structural compliance and strong surface adhesion, allowing integration into stretchable and bendable devices without sacrificing performance. [30] Beyond nanowires, carbon-based materials, particularly graphene, have gained significant attention as electrode candidates due to their high conductivity, mechanical robustness, and tunable flexibility. Graphene's atomic thinness, high carrier mobility, and outstanding tensile strength provide superior flexibility and fatigue resistance compared to traditional thin-film materials. [31,32] Three-dimensional porous graphene architectures are especially promising, as their large surface area and interconnected networks enhance charge transfer, increase electrode‑polymer interaction, and maintain mechanical stability under deformation. [33] Laser-induced graphene (LIG) offers a particularly compelling route for realizing such architectures in flexible devices. This direct-write method photothermally converts polymer surfaces into a porous, conductive graphene network, enabling in-situ patterning without additional transfer steps. [34] LIG is scalable, adaptable to diverse substrate geometries, and has demonstrated strong performance across numerous applications, [35–37] including photodetectors, [38] bioelectronics, [39] energy storage devices, [40] and chemical sensors. [41] Despite these advances, the use of LIG and other graphene-based porous conductors as electrodes in ultrasound transducers remains largely unexplored. Their inherent mechanical compliance, high conductivity, and process versatility present an opportunity to address long-standing interfacial mismatch challenges in flexible ultrasound transducers while potentially enhancing piezoelectric and acoustic performance. In our previous work [42], we demonstrated flexible ultrasound transducers based on LIG electrodes integrated with the additive manufacturing of poly(vinylidene fluoride-trifluoroethylene) (PVDF-TrFE) ink, highlighting the advantages of porous graphene architectures for strong interfacial bonding and improved acoustic performance. A schematic illustration of the fabrication process, integrating the photothermal laser process for LIG formation with additive manufacturing of drop-casted piezopolymer ink, is shown in **Figure 1a**.

Building on our earlier work with LIG-integrated piezopolymer transducers (**Figure 1b**), this study shifts the emphasis from piezopolymer processing to the role of the bottom-electrode interface. Here, we examine how differences in electrode composition and morphology, ranging from dense metallic films to layered graphene flakes and porous laser-induced graphene, affect piezoelectric response, dielectric behavior, and acoustic output in flexible PVDF-TrFE–based transducers. By comparing Ag, Au, GF, LIG, and Au-decorated LIG electrodes under identical processing and poling conditions, we directly examine the influence of interfacial structure on mechanical compliance, charge-transfer efficiency, and overall electromechanical performance of FUSTs. The results provide clear evidence of how porous, compliant electrodes promote stronger coupling and improved stability, while dense or weakly bonded interfaces limit performance and durability. Collectively, these findings establish electrode–piezopolymer interface engineering as a key design consideration for flexible ultrasound transducers and point to LIG-based architectures as a practical route toward high-performance, conformable systems for wearable and diagnostic imaging applications.

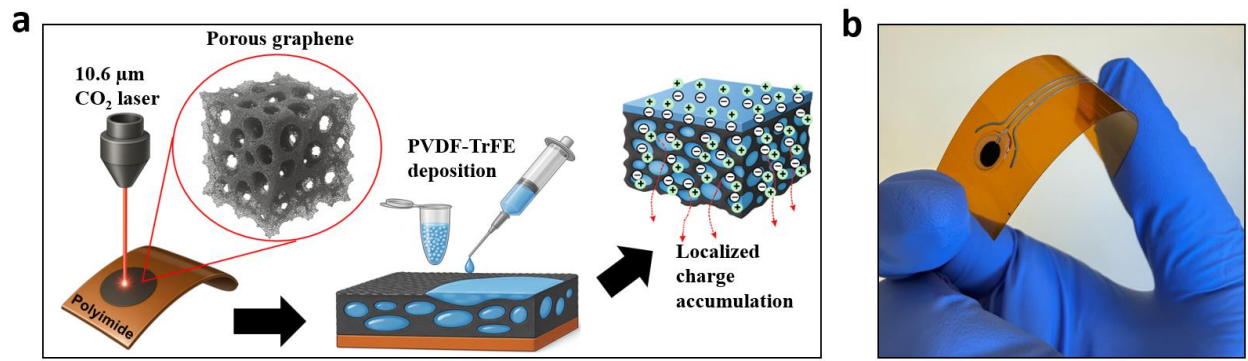

**FIGURE 1.** Fabrication and structure of the LIG/PVDF-TrFE composite. (a) Schematic depicting porous LIG formation, piezopolymer deposition, and enhanced interfacial charge accumulation enabled by the LIG architecture. (b) Photograph of the flexible LIG/PVDF-TrFE ultrasound transducer.

## 2. Results and discussion

### 2.1 Fabrication of piezopolymer-based flexible transducers

To investigate the influence of electrode composition and morphology on the electromechanical behavior of flexible ultrasound transducers (FUSTs), devices were fabricated on 125 μm-thick polyimide (PI) substrates using various bottom-electrode materials (**Figure 2a**), including conventional metals (Ag, Au), graphene flakes (GF), and LIG. The fabrication processes were tailored for each material to produce distinct electrode–piezopolymer interfaces (Experimental Section). Cross-sectional SEM images reveal clear differences in interfacial morphology and layer thickness across the electrode types (**Figures 2b–e**). LIG forms a porous, interconnected layer with a thickness of $25 \pm 3$ μm directly contacting the uniform PVDF-TrFE film ($60 \pm 5$ μm), whereas silver is dense and smooth ($98 \pm 5$ μm), and GF exhibits a stacked, heterogeneous structure ($20 \pm 2$ μm). The porous LIG network enables partial PVDF-TrFE infiltration and mechanical interlocking, which increases surface interaction of the electrode and piezopolymer, and improves strain transfer at the interface. In contrast, the dense Ag layer and the flake-stacked GF interface provide limited mechanical coupling and fewer charge-transfer pathways through the interface plane. Cross-sectional SEM images of bare GF electrodes (**Figure S1**) further confirm their layered, heterogeneous nature, consistent with their limited interfacial contact with the piezopolymer. These morphological variations are critical, as the interfacial structure governs both mechanical compliance and charge-transfer efficiency, together dictating the overall electromechanical response and long-term durability of the FUSTs. To further tailor the interfacial properties, gold nanoparticles were incorporated onto the LIG surface to form nano-gold-decorated LIG (LIG-Au) electrodes (Experimental Section). SEM images of the LIG-Au electrode (**Figure S2**) confirm the uniform decoration of Au nanoparticles across the graphene network without compromising its porous framework.

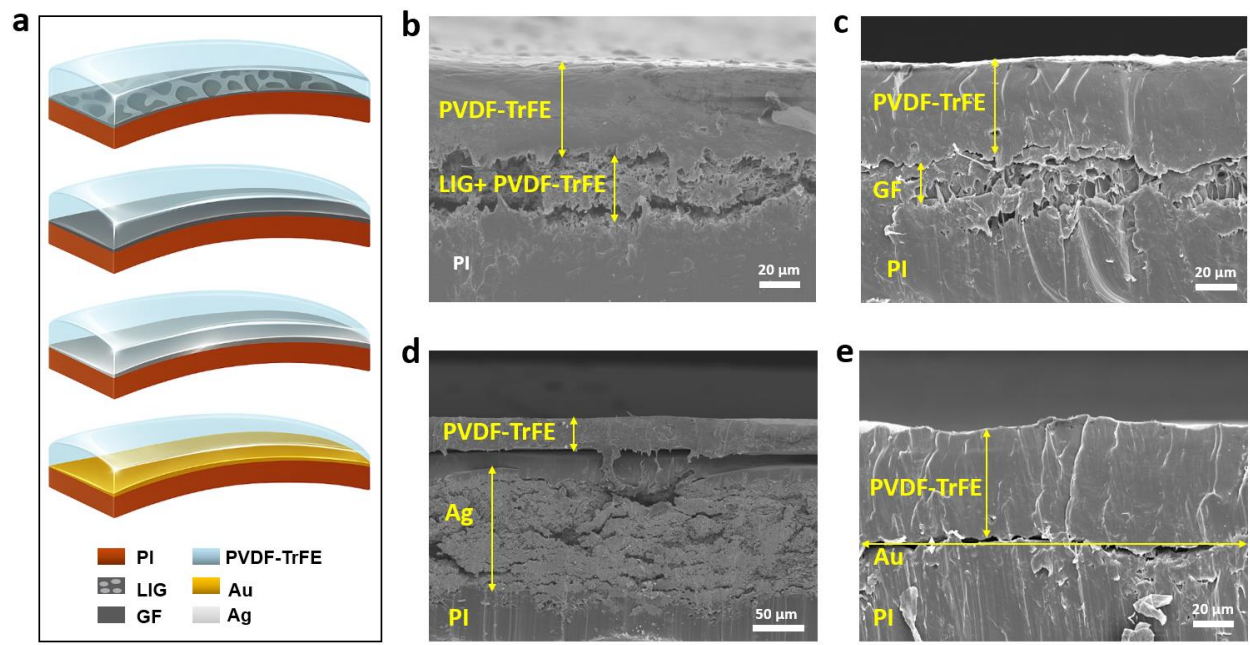

**FIGURE 2.** Structural configurations and electrode–polymer interfaces of the fabricated transducers. (a) Schematic illustration of the multilayer structures for piezopolymer-based ultrasound transducers incorporating LIG, graphene flakes (GF), silver (Ag), and gold (Au) bottom electrodes. (b–e) Cross-sectional SEM images of piezopolymer layers deposited on (b)

LIG, (c) GF, (d) Ag, and (e) Au electrodes, showing the distinct interfacial morphologies and layer thicknesses associated with each electrode type.

## 2.2 Effect of poling electric field, temperature, and time on PVDF-TrFE polarization

The poling process strongly influences dipole orientation and the resulting piezoelectric activity of PVDF-TrFE. To identify the optimal parameters, the applied field strength, temperature, and duration were varied. The corona-poling configuration used for dipole alignment within the piezopolymer matrix in the LIG-based transducer is shown in **Figure 3a**. The piezoelectric charge coefficient ($d_{33}$) of devices poled at different conditions was measured using the Berlincourt test setup (**Figure S3**). **Figure 3b** presents $d_{33}$ as a function of poling field from 140 to 290 V/μm (n=4 devices per group), while the poling temperature and duration were kept constant at 80 °C and 60 min, respectively. At 140 V/μm, $d_{33}$ was 4.8 ± 1.2 pC/N, increasing to a maximum of 23.1 ± 0.5 pC/N at 170 V/μm and remaining nearly constant thereafter, indicating saturation of dipole alignment. It should be noted that excessively high electric fields can damage the piezopolymer surface, forming microscopic defects or pinholes that promote leakage and degrade device-level performance. Because the Berlincourt test provides a localized, quasi-static measure of piezoelectric response over the probe contact area, dynamic compression tests were also performed to evaluate the device-level electromechanical output under operating conditions. **Figures 3c–3e** show the effect of field strength, temperature, and poling duration on voltage amplitude during cyclic compression (55 kHz, 4 N). The output voltage rose with increasing field and reached a maximum of 20.1 ± 0.2 mV when poled at 200 V μm$^{-1}$, 80 °C, and 60 min. Temperatures near 80 °C yielded the strongest signal, likely due to enhanced chain mobility. [43] Similarly, the dependence on poling duration revealed a pronounced optimum at 60 min (**Figure 3e**).

Electrical characterization was also performed to evaluate the impedance, capacitance, and relative permittivity under operating conditions (**Figure S4, S5, and S6**). These results establish that a poling field of 200 V µm⁻¹, temperature of 80 °C, and duration of 60 min provide an optimal balance between dipole alignment and film integrity. These parameters were therefore adopted for all subsequent device fabrication and characterization.

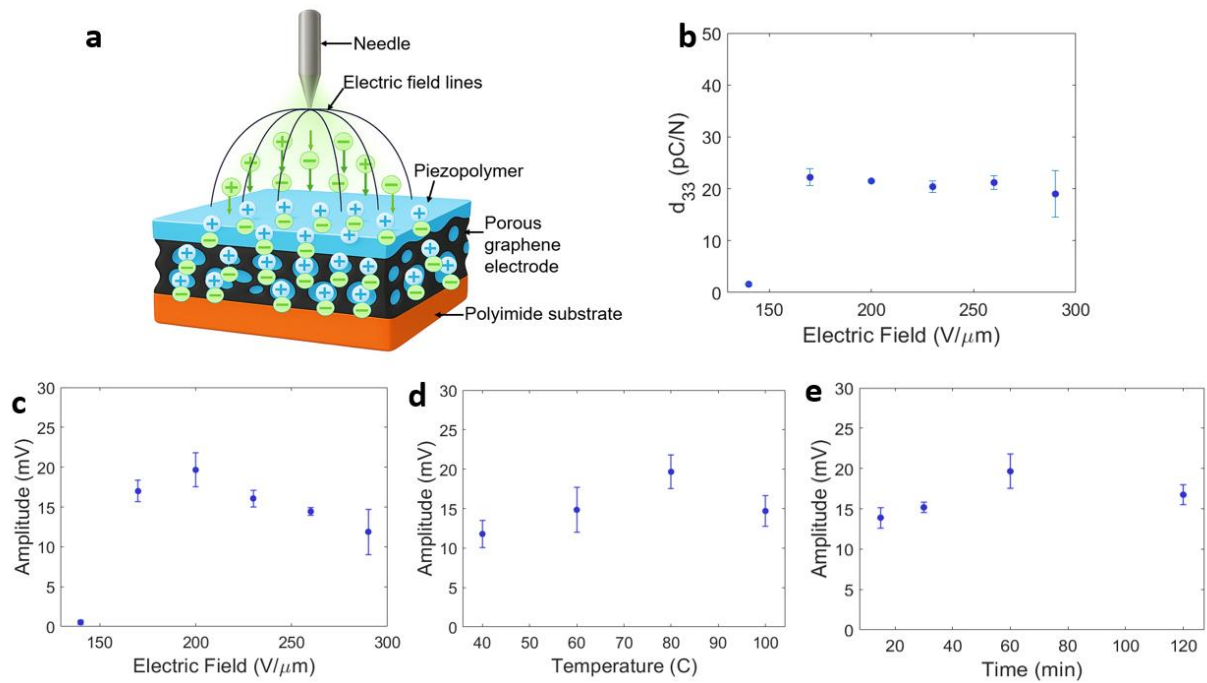

**FIGURE 3.** Corona poling optimization of piezopolymer. (a) Schematic of the corona-poling configuration used to align β-phase dipoles within the LIG/PVDF-TrFE composite. (b) Piezoelectric charge coefficient ($d_{33}$) as a function of applied poling field. (c–e) Output voltage amplitudes under dynamic compression as a function of (c) poling field, (d) poling temperature, and (e) poling duration.

## 2.3 Dielectric, electrical, and piezoelectric properties of piezopolymer-based transducers with different bottom electrodes

The electrical impedance, dielectric, and piezoelectric properties of piezopolymer-based transducers with varying bottom electrodes were compared to explain the influence of electrode composition and interfacial structure on their electromechanical performance (**Figures 4a–f**).

The impedance displays the characteristic capacitive behavior of piezoelectric composites, decreasing with frequency for all devices. Among the tested electrodes, LIG–Au exhibited the lowest impedance across the measured range, reflecting its enhanced electrical conductivity due to Au nanoparticle decoration and the interconnected pathways within the porous LIG framework. In contrast, Ag electrodes showed the highest impedance, while Au, LIG, and GF devices fell between these extremes depending on their interfacial continuity and surface roughness (**Figure 4a**). Areal capacitance data reinforced these observations, as LIG-Au exhibited capacitance values greater than those of Ag, indicating that a porous and continuous electrode–polymer interface promotes greater charge accumulation and interfacial polarization (**Figure 4b**). Correspondingly, dielectric constant measurements revealed that the LIG-Au transducer reached values of approximately 14 at low frequency, stabilizing near 12 at higher frequencies, which aligns with increased electroactive surface area and localized dipole polarization. LIG demonstrated superior dielectric behavior compared with metallic electrodes, confirming that morphological porosity alone substantially enhances interfacial capacitance even without metallic nanoparticles (**Figure 4c**). Collectively, these findings indicate that while electrode conductivity determines the overall impedance behavior, the interplay between electrical continuity and structural heterogeneity governs dielectric enhancement. Dense metallic Ag films limit interfacial polarization pathways, whereas porous graphitic morphologies enhance surface interaction with the piezopolymer, enabling greater charge accumulation within the active layer.

To evaluate the piezoelectric properties of the piezopolymer in transducers, we measured the output voltage of the transducers under sinusoidal compression, the piezoelectric coupling coefficient ($d_{33}$), and the piezoelectric voltage constant ($g_{33}$). **Figure 4d** shows that the Au-based transducer exhibited the highest voltage amplitude of ~25 mV, closely followed by the LIG-, and LIG-Au-based devices. In contrast, the GF- and Ag-based transducers generated noticeably weaker signals of ~10 mV. Consistent with this trend, LIG, LIG-Au, and Au-based transducers show larger $d_{33}$ (~25 pC/N) than GF- and Ag-based transducers (~21 and 18 pC $N^{-1}$, respectively) (**Figure 4e)**. These results could potentially indicate that both Au and LIG interfaces promote efficient electromechanical coupling through different mechanisms: the continuous, thin Au surface ensures uniform charge collection, whereas the porous, compliant LIG structure facilitates improved stress transfer and local field distribution within the piezopolymer layer.

The voltage sensitivity, described by the piezoelectric voltage constant $g_{33}$, provides additional context for interpreting these results. Because $g_{33}$ depends on the ratio of $d_{33}$ to the permittivity ($g_{33}=d_{33}\,\varepsilon^{-1}$), materials with lower permittivity tend to produce larger voltage outputs for the same applied stress. As shown in **Figure 4f**, Au exhibits the highest $g_{33}$ (~0.30 V m $N^{-1}$), owing to its combination of a relatively large $d_{33}$ and a lower permittivity, which efficiently converts applied stress into voltage. Ag follows (~0.25 V m $N^{-1}$), where its lower $d_{33}$ is partly compensated by an even smaller dielectric constant, resulting in a higher $g_{33}$ than expected from charge generation alone. LIG, LIG–Au, and GF fall within a narrower range (~0.18–0.20 V m $N^{-1}$), reflecting their higher effective permittivity. This increased permittivity suppresses the voltage output, despite improved mechanical coupling and, in the case of LIG, strong $d_{33}$ values. These results show that $d_{33}$ and $g_{33}$ highlight different aspects of electromechanical behavior. Au exhibits both a high $d_{33}$ and the highest $g_{33}$, indicating efficient conversion of mechanical stress into electrical output. LIG also shows a strong $d_{33}$, and although its $g_{33}$ is slightly lower due to its higher effective permittivity, its porous and compliant interface promotes more effective stress transfer within the piezopolymer layer. In contrast, Ag exhibits the lowest $d_{33}$ and a slightly rigid, nonporous interface, which limits mechanical coupling despite its moderate $g_{33}$. These results demonstrate that electromechanical performance is governed not only by intrinsic piezoelectric coefficients of the piezoelectric material but also by the ability of the electrode–polymer interface to transmit stress and support charge development.

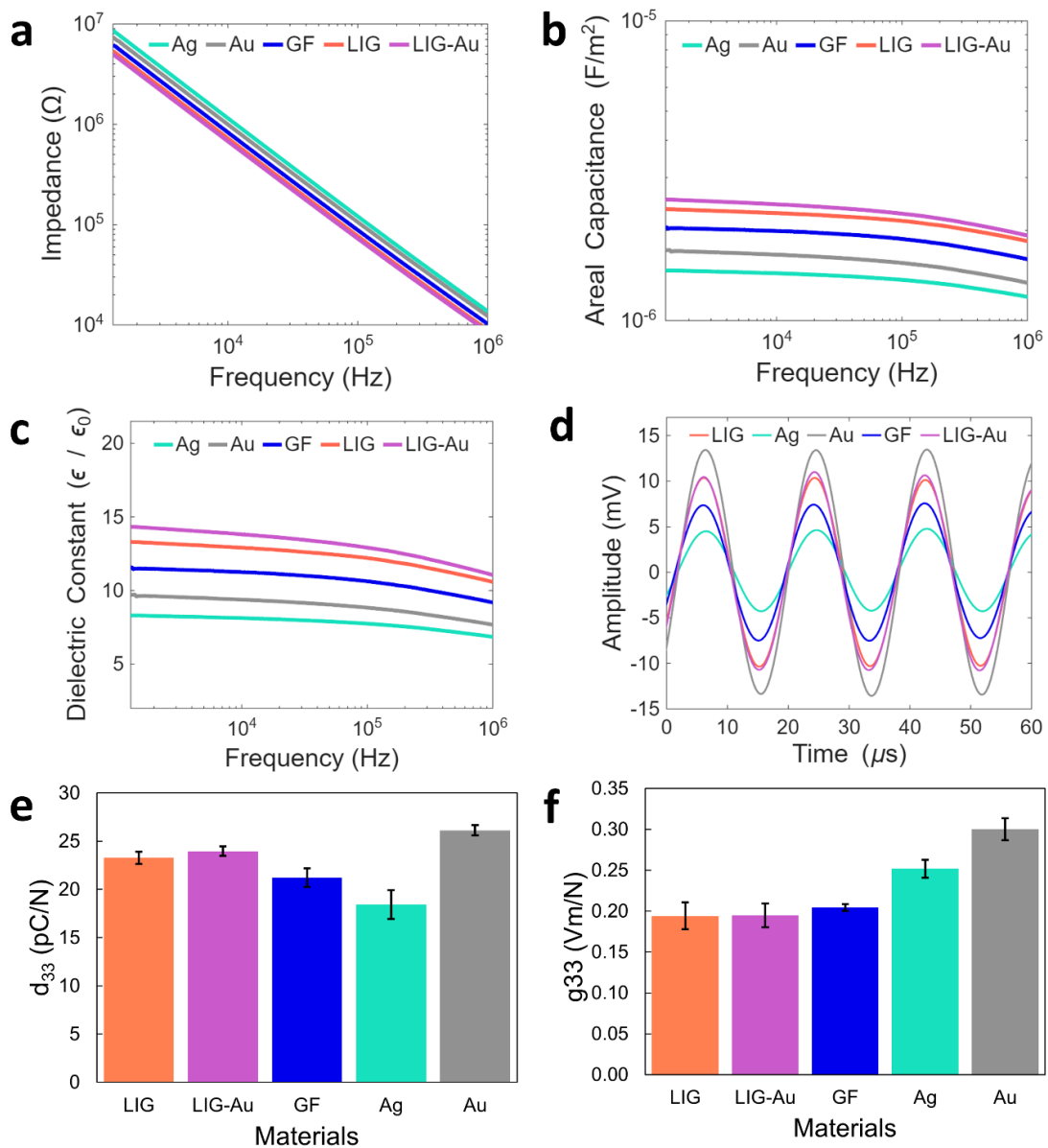

**FIGURE 4.** Dielectric, electrical, and piezoelectric properties characterization of piezopolymer-based transducers with different bottom electrodes. (a) Impedance (b) Areal capacitance, and (c)

Dielectric constant of PVDF-TrFE film deposited on different bottom electrodes: Ag, Au, GF, LIG, and LIG-Au. (d) Voltage response under cyclic compression, (e) $d_{33}$, and (f) $g_{33}$ of piezopolymer-based transducers with different bottom electrodes.

## 2.4 Acoustic output of piezopolymer-based transducers with different bottom electrodes

To investigate how interfacial properties influence device performance in transmission and reception, we conducted pulse–echo measurements on the fabricated FUTs. This technique directly assesses transducer function by measuring key parameters such as echo amplitude and bandwidth [44], offering clear insights into the relationship between electrode–polymer morphology and ultrasound emission. The pulse-echo tests were conducted in a water tank as explained in the Experimental Section. After obtaining the time-domain signal, the data were converted to the frequency domain using the Fast Fourier Transform (FFT) **(Figure 5a-e)**. The resulting pulse–echo responses revealed clear differences in signal amplitudes and frequency characteristics among FUSTs fabricated using different electrode materials. LIG-based transducers produced the largest peak-to-peak amplitude of about 6 V at a central frequency of ~18 MHz, which is nearly twice that of Ag- and GF-based devices (**Figure 5f, g**). This enhanced output can be attributed to the porous LIG architecture, which allows partial infiltration of PVDF-TrFE and increased interfacial contact, thereby supporting more efficient stress transfer and electromechanical coupling. Au-based devices provide signal amplitudes comparable to those of LIG-based devices (**Figure 5f**), likely due to their uniform, thin films with strong conductive continuity, which facilitates efficient charge collection. LIG-Au devices exhibited a slightly reduced amplitude relative to pristine LIG despite similar $d_{33}$ and $g_{33}$ values. This decrease could potentially be attributed to nanoscale stiffening and mass loading introduced by Au nanoparticles within the porous LIG matrix, which alter local impedance and redistribute acoustic energy across several closely spaced resonant modes. As a result, a portion of the emitted energy is stored in secondary oscillations rather than contributing to the main echo amplitude, consistent with the more pronounced ringing observed in the time–frequency response (**Figure 5b**).

To evaluate how electrode composition influences the resonant behavior of the FUTs, we analyzed the quality factor (Q-factor), which characterizes the ratio of acoustic energy stored in the transducer to that lost per cycle.[45] Silver and GF electrodes exhibited the lowest Q-factors (2.3 and 2.6), indicating strongly damped responses with minimal ringing (**Figure 5h**). LIG and Au showed intermediate Q-values (4.0 and 4.8), while LIG-Au produced the highest Q-factor (6.1), consistent with its extended ringing and multiple spectral components in the time–frequency response. Bandwidth trends followed this behavior (**Figure 5i**). Because Q is inversely related to bandwidth (Q ~ CF/BW), GF and Ag, with the lowest Q-factors, produced the broadest bandwidths (~8 MHz). In contrast, LIG, Au, and LIG-Au showed narrower bandwidths, matching their higher Q-values. Bandwidth is defined as the frequency range over which a device operates effectively, and it influences image resolution [46] Typically, a broader bandwidth enhances axial resolution, which is essential for resolving fine anatomical features. [47] Our findings show that electrode selection shifts the balance among echo amplitude, bandwidth, and resonance characteristics. LIG provides the highest echo amplitude, whereas LIG-Au exhibits the strongest resonance and the tightest narrowband response. This demonstrates that FUT performance can be tuned to the needs of specific imaging applications.

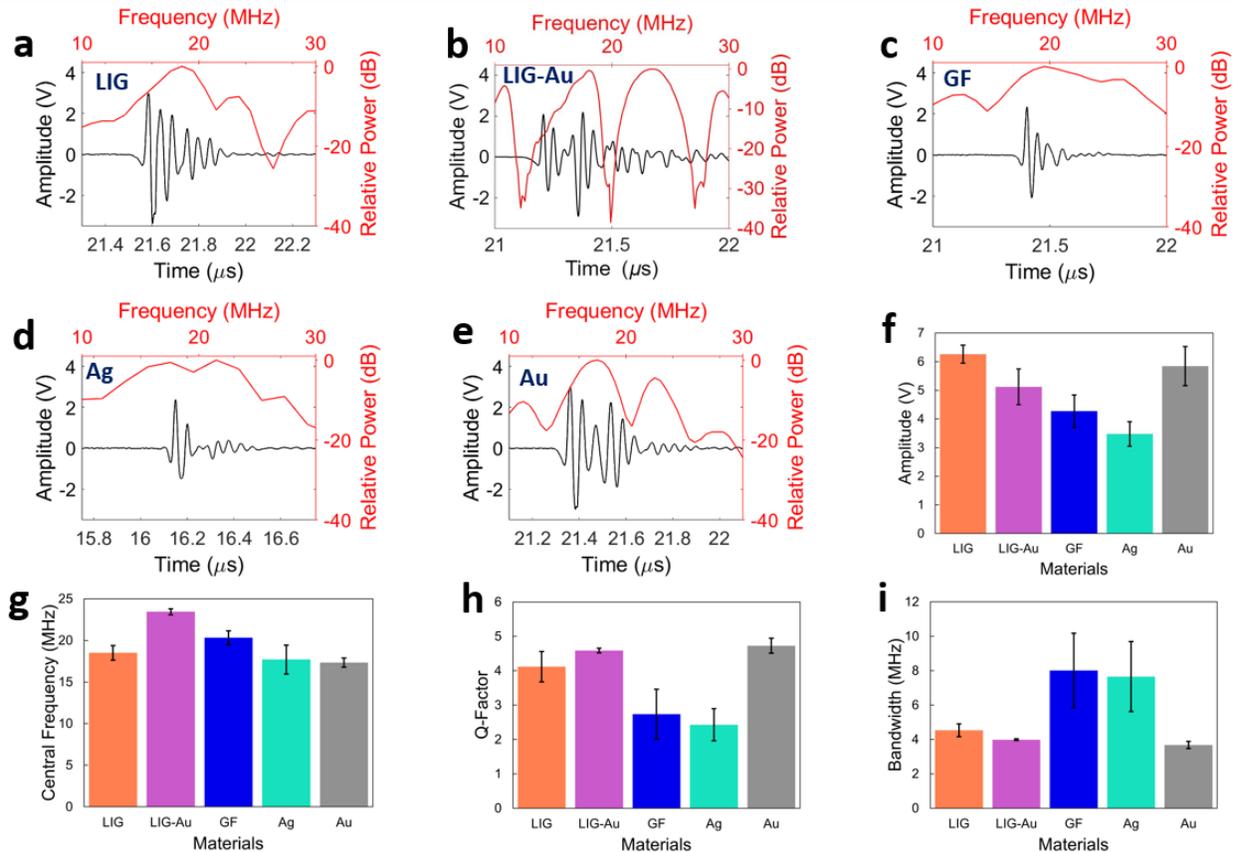

**FIGURE 5.** Acoustic performance of piezopolymer-based transducers. Pulse-echo signal and the calculated frequency response of ultrasound transducers with (a) LIG, (b) LIG-Au, (c) GF, (d) Ag, and (e) Au bottom electrodes. Comparison of the (g) central frequency, (h) Q-Factor, and (i) Bandwidths of the ultrasound transducers with 100 µm PVDF-TrFE thickness and different bottom electrodes.

## 2.5 Characterizations of the mechanical robustness, flexibility, and durability of piezopolymer-based transducers with different bottom electrodes

To evaluate the suitability of the transducers for flexible and wearable ultrasound applications, we characterized their mechanical robustness and long-term stability under repeated bending and continuous operation (**Figure S7, S8, and Movie S1**). These assessments are essential because mechanical fatigue, interfacial degradation, and electromechanical drift commonly limit the reliability of flexible piezoelectric devices. A bending setup (**Figure 6a**) was used to apply cyclic deformation corresponding to a 22.6 mm bending diameter (**Video S1**), and pulse–echo measurements were collected throughout 10,000 bending cycles (**Figure S7**). The LIG-based transducer maintained a stable output amplitude of ~6 V throughout testing, demonstrating strong resistance to strain-induced degradation (**Figure 6b**). This stability is attributed to the compliant, porous LIG structure, which distributes mechanical strain and preserves adhesion at the electrode–piezopolymer interface. GF-based devices exhibited moderate stability but began to decline after approximately 1,000 cycles, likely due to sliding and microcracking within the stacked flake structure (**Figure 6b**). In contrast, metallic electrodes showed more pronounced degradation (**Figure 6b**): Ag-based devices gradually decreased in amplitude over multiple thousands of bending cycles, while Au-based devices exhibited sharp attenuation, losing nearly 80% of their signal, after only a few hundred cycles, reflecting their limited ability to accommodate bending without interfacial failure. It should be noted that PI substrates used for GF-, Ag-, and Au-based transducers were laser-texturized (Experimental Section) to enhance adhesion between the electrode, piezopolymer layer, and substrate. Without this texturization, Ag-based transducers experienced complete mechanical failure before reaching 10,000 bending cycles. [42] Longevity tests were performed using pulse–echo measurements over an 8-week period (**Figure 6c** and **Figure S8**). Among all devices, only the Ag-based transducers showed a steady decline in signal amplitude after weak 4. Because the LIG-based transducers demonstrated high flexibility and strong long-term stability, we further conducted continuous pulse-echo measurements for 8 hours, yielding a consistent signal output (**Figure 6d**), indicating the potential of these transducers for continuous and longitudinal monitoring. Therefore, our LIG-based piezoelectric transducer is promising for use in flexible and wearable ultrasound applications and longitudinal monitoring.

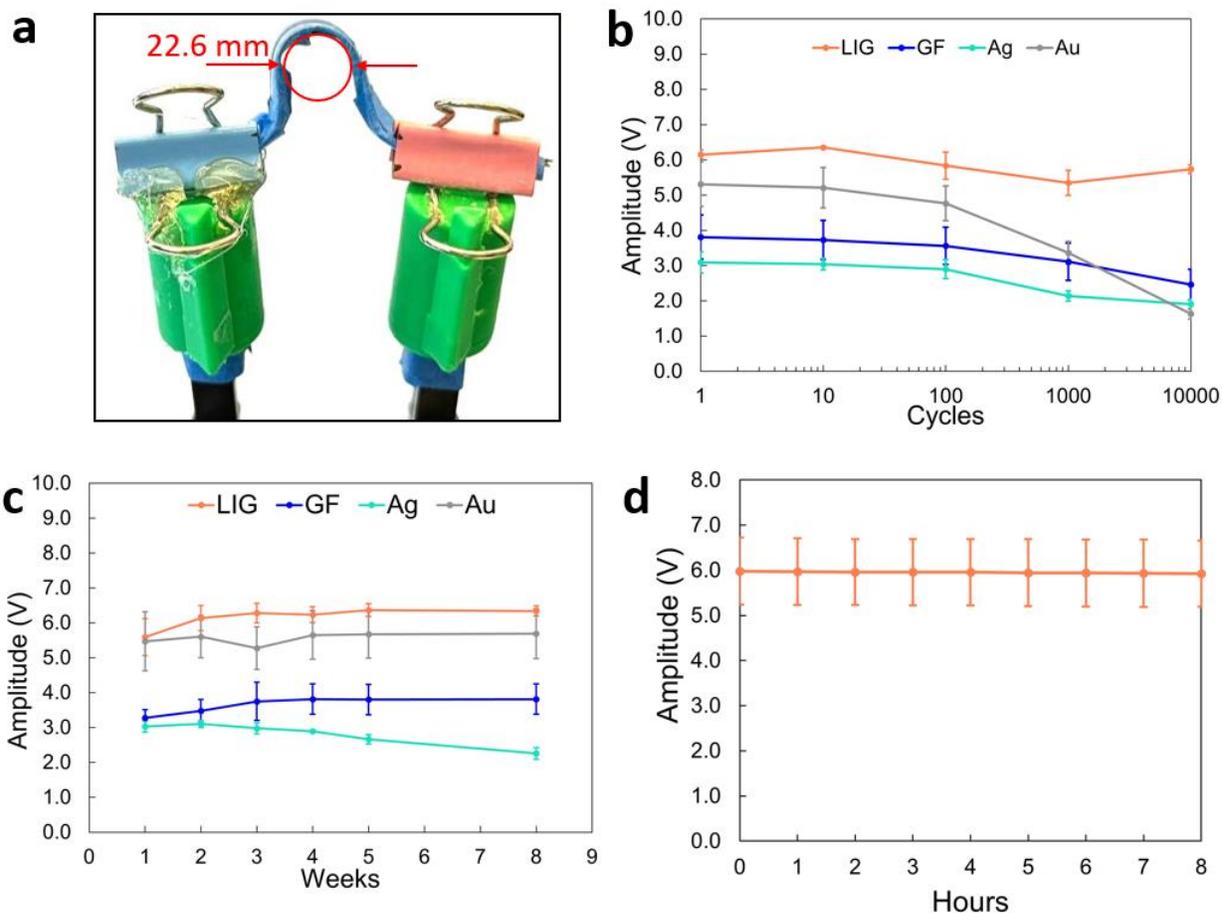

**FIGURE 6.** Mechanical robustness and long-term stability of piezopolymer-based transducers with different bottom electrodes. (a) Photograph of the bending setup used for cyclic mechanical testing (bending radius: 22.6 mm). (b) Output amplitude of LIG-, GF-, Ag-, and Au-based transducers during cyclic bending up to 10,000 cycles, showing the superior mechanical endurance of the LIG configuration. (c) Long-term aging test over 8 weeks, demonstrating stable signal retention for LIG-, GF-, and Au-based transducers, with progressive degradation observed in Ag-based transducers. (d) Continuous pulse–echo operation of the LIG-based transducer over 8 hours, exhibiting highly stable acoustic output suitable for continuous and longitudinal monitoring.

## 3. Conclusion

In summary, this study examined how different bottom-electrode materials influence the electromechanical performance of flexible piezopolymer-based ultrasound transducers by modifying the structure and quality of the electrode–piezopolymer interface. Using Ag, Au, GF, LIG, and LIG-Au electrodes, we created transducers with distinct interfacial morphologies and compared their dielectric property, impedance behavior, piezoelectric output, and acoustic performance. Among the tested materials, LIG formed the most effective interface for piezopolymer coupling: its porous architecture enabled partial PVDF-TrFE infiltration, enlarged the interfacial contact area, and supported efficient stress transfer, leading to strong piezoelectric response and acoustic outputs. Au electrodes, despite lacking infiltration, produced comparable device-level acoustic amplitudes due to their uniform, continuous, and highly conductive surface that facilitates efficient charge collection. LIG-Au electrodes showed similar $d_{33}$ and $g_{33}$ values to LIG but slightly reduced acoustic output, could potentially reflecting subtle changes in local mechanical and acoustic impedance introduced by nanoscale Au decoration. In contrast, GF and Ag electrodes, with limited interfacial compliance and fewer charge-transfer pathways, produced noticeably weaker electromechanical and acoustic responses. However, the relatively thick electrode interface (as verified by SEM images) may have led to increased dampening resulting in a more broadband (low Q) acoustic response. Mechanical and aging tests further underscored the influence of interface structure. LIG-based devices retained consistent performance after 10,000 bending cycles and throughout an eight-week longevity study, whereas Ag, Au, and GF devices exhibited varying degrees of degradation under bending, and Ag showed a measurable decline over time. These trends highlight that interfacial compliance and bonding, not simply electrode conductivity, are central to achieving durable and high-performing flexible transducers. The results demonstrate that controlling electrode morphology provides a practical strategy for improving the efficiency, stability, and flexibility of piezopolymer-based ultrasound transducers. LIG, in particular, offers a compelling combination of mechanical flexibility, strong electromechanical coupling, and robust acoustic output, positioning it as a promising electrode material for future wearable and conformal ultrasound systems.

## 4. Experimental Section

### 4.1 Fabrication of LIG bottom electrode

LIG was synthesized from commercial polyimide (PI) film (125 μm-thick) (18-5F-FPC, CS Hyde) using a photothermal laser treatment to form a circular working electrode with a diameter of 3.5 mm. A 10.61 μm $CO_2$ laser engraving and cutting system (Universal Laser Systems, VLS2.30) was operated at a power of 4.2 W, scan speed of 3.5 in s$^{-1}$, pulse per inch of 1000 (PPI), and pulse duration of 14 μs under ambient conditions to pattern the LIG.

### 4.2 Fabrication of LIG-Au bottom electrode

LIG–Au electrodes were fabricated by introducing a gold precursor into the porous LIG network, followed by a secondary laser irradiation to immobilize gold nanoparticles within the LIG matrix. [48] A 0.5 mol L−1 $H_2SO_4$ (sulfuric acid) solution was prepared by diluting $H_2SO_4$ (ACS, 95.0%-98.0%, Thermo Fisher Scientific) with deionized water. $HAuCl_4 \cdot 3H_2O$ ('Tetrachloroauric(III) acid trihydrate, ACS reagent', Thermo Fisher Scientific) was then dissolved in the $H_2SO_4$ solution to a 0.1 M $HAuCl_4$ precursor solution. [49] The precursor solution was drop-cast onto LIG electrodes fabricated using a 10.6 μm $CO_2$ laser system at a power of 3.6 W, PPI of 1000, and scan speed of 3.5 in s$^{-1}$. After drying, a second laser irradiation was performed at 1.2 W, PPI of 1000, and a scan speed of 1.75 in s$^{-1}$ to anchor gold nanoparticles within the LIG structure.

### 4.3 Fabrication of the graphene flake, silver, and gold bottom electrodes

To mitigate delamination in Ag-, Au-, and GF-based transducers, the PI substrate surrounding the electrode area was laser-textured over a 7.5 mm-diameter region using a $CO_2$ laser (Fusion Maker 12, Epilog Laser) operated at 8 W, 1200 dpi, and a scan speed of 29.5 in $s^{-1}$.

For GF-based devices, an additional 3.5 mm-diameter well was engraved at the electrode area using the same laser parameters. This well-defined the active electrode region and confined the drop-casting of a graphene flake dispersion in water (flake size 1–3 µm, ACS Material), ensuring consistent electrode dimensions. After deposition, GF devices were dried under vacuum to remove residual solvent and trapped gas that could disrupt flake packing.

For Ag-based devices, the 3.5 mm-diameter electrode pad was printed simultaneously with the interconnects using conductive silver ink (FS0142, ACI Materials). For Au-based devices, a gold electrode layer with a thickness of approximately 80 nm was deposited by ion-beam sputtering (SPT-20, Coxem) at a current of 3 mA for three sequential 300 s intervals. A shadow mask was used during deposition to maintain a constant electrode diameter of 3.5 mm.

For gold-based devices, a gold bottom electrode with a thickness of approximately 80 nm was deposited by an ion sputter coater (SPT-20, Coxem) at a current of 3 mA for three sequential 300 s intervals. A shadow mask was used during deposition to maintain a constant electrode diameter of 3.5 mm.

**4.4 Fabrication of piezopolymer-based ultrasound transducers**

Electrical contact pads and interconnects were printed using a direct-write extrusion system (Voltera, NOVA) with conductive silver ink (FS0142, ACI Materials) after fabrication of bottom electrodes as described above. The printed devices were annealed at 150 °C for 15 min to cure the ink. After curing, a 6.5 mm-diameter ring of UV-curable silicone (SS-5083 UV Dual Cure Acetoxy Silicone, Silicone Solutions) was deposited around the active area of each device and cured under UV illumination, followed by thermal curing at 60 °C for 20 min. PVDF-TrFE ink (15 mg; Piezotech FC20 ink, Arkema) was then drop-cast into the silicone well to form a piezoelectric layer with a thickness of approximately 60 μm, followed by degassing under vacuum for 1 h. The PVDF-TrFE layer was dried and annealed at 140 °C for 20 min. The transducers were subsequently subjected to corona-discharge poling by placing them on a hot plate beneath a needle-type corona electrode array and heating to 80 °C. A DC electric field of 170 V μm$^{-1}$ was applied for 60 min, after which the electric field was maintained while the temperature was gradually reduced to 40 °C over 40 min. A gold top electrode ($\sim$60 nm thick) was then deposited by ion sputter coater (SPT-20, Coxem) at a current of 3 mA for 300 seconds. Finally, a 5 μm thick parylene (Parylene-C dimer) layer was deposited by chemical vapor deposition using a commercial parylene coating system (Labcoter 2, SCS) to encapsulate and passivate the device.

## 4.5 Dielectric and Piezoelectric Characterization

The electrical impedance, capacitance, and permittivity of the transducers were measured using an LCR meter (IM3536, HIOKI). The piezoelectric coefficients of the transducers were measured using a $d_{33}$ meter (Piezo d33 Test System, ACP International) at a load of 3 N, measured using a static force sensor (PK-D3-F10N, PolyK).

Compression testing of the devices was conducted using the same $d_{33}$ setup, replacing the $d_{33}$ probe with an actuator (PK2FSP2 - Discrete Piezo Stack, Thorlabs). Additionally, a 10 MHz function/arbitrary waveform generator (33210A, Agilent) and a higher fidelity oscilloscope (HDO4024A, Teledyne Lecroy) that could measure in 10 GS/s intervals were used to capture the data. Samples were tested at 6 N of load and 55 kHz excitation between 0 and 10 V.

## 4.5 Acoustic Output Measurement via Pulse-Echo

Pulse-echo measurements were performed via a water tank setup, consisting of a Plexiglas tank and an aluminum disk mounted on a motorized positioning rack inside of the water tank for reflection. A pulser/receiver (5073PR, OLYMPUS) was used to obtain the signal amplitude, central frequency, and bandwidth of the devices, which were mounted on the side of the tank with conductive gel to eliminate any air gaps between them and the barrier into the water. The pulser/receiver was set at a transmit energy of level 4, damping level of 1, pulse repetition frequency (PRF) of 200 Hz, and receiver gain of 30 dB. After mounting the device and turning on the pulser/receiver, the distance and angle of reflection of the aluminum disk was adjusted to maximize the signal amplitude. The two-way echo response was then directly captured by the same device pulsing and visualized on an oscilloscope (T3DSO2204A, Teledyne LeCroy). The frequency domain was then analyzed via fast Fourier transform (FFT). The center frequency (CF) and $-6$ dB bandwidth (BW(%)) were determined via the following equations:

$$CF = \frac{f_1 + f_2}{2} \qquad\qquad 1$$

$$BW(\%) = \frac{f_2 - f_1}{CF} \times 100\% \qquad\qquad 2$$

Where $f_1$ and $f_2$ represent the frequencies where the FFT magnitude of the echo drops by $-6$ dB, with $f_1$ being lower than $f_2$.

## 4.6 Bending Tests

Cyclic bending tests were performed using a programmable robotic arm (Yahboom DOFBOT AI Vision Robotic Arm). A custom fixture was designed to hold each device between two grippers such that the active electrode region was positioned at the apex of the bending curvature. The robotic arm was programmed to bend the device to a radius of 22.6 mm over 0.4 s, hold this position for 0.5 s, and then return to a flat configuration over 0.4 s. Devices were removed and characterized by pulse–echo measurements after 1, 10, 100, 1,000, and 10,000 bending cycles.

**Supporting Information**

Supporting Information is available from the Wiley Online Library or from the author.

**Acknowledgements**

Effort sponsored by the Government under Other Transaction Number W81XWH-15-9-0001. The authors acknowledge the support of Dr. Pilgyu Kang for access to the $CO_2$ laser engraving and cutting system (Universal Laser Systems, VLS2.30) used in the fabrication of the LIG electrodes.

**Disclaimer**

The views and conclusions contained herein are those of the authors and should not be interpreted as necessarily representing the official policies or endorsements, either expressed or implied, of the U.S. Government.

**Conflict of Interest**

The authors declare no conflict of interest.

**Data Availability Statement**

The data that support the findings of this study are available from the corresponding author upon reasonable request.

Flexible piezopolymer-based ultrasound transducers are engineered by tailoring the electrode–piezopolymer interface across metallic, flake-based, and porous graphene electrodes. The porous architecture of laser-induced graphene enables partial polymer infiltration, improving interfacial coupling, piezoelectric response, and acoustic output compared with dense and weakly bonded electrodes. LIG-based devices maintain stable performance under long-term bending and aging, demonstrating superior mechanical durability. This interface-engineering strategy identifies LIG as an effective electrode for high-performance, wearable ultrasound systems.

Spencer Hagen[§], Dulcce A Valenzuela[§], Parag V Chitnis *, Shirin Movaghgharnezhad *

**Microstructured Electrode-Piezopolymer Interface for Ultrasound Transducers with Enhanced Flexibility and Acoustic Performance**

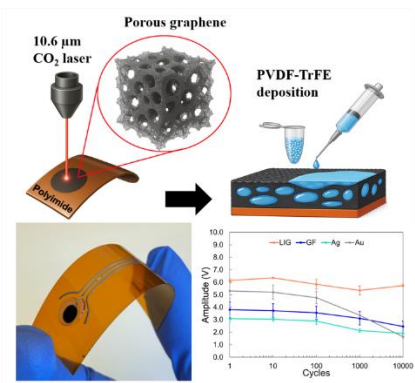

**Supporting Information**

**Microstructured ElectrodePiezopolymer Interface for Ultrasound Transducers with Enhanced Flexibility and Acoustic Performance**

Spencer Hagen[§], Dulcce A Valenzuela[§], Parag V Chitnis *, Shirin Movaghgharnezhad *

[§]These authors contributed equally to this work.

**List of contents:**

1. Supplementary Figures

2. Supplementary Movie

# 1. Supplementary Figures

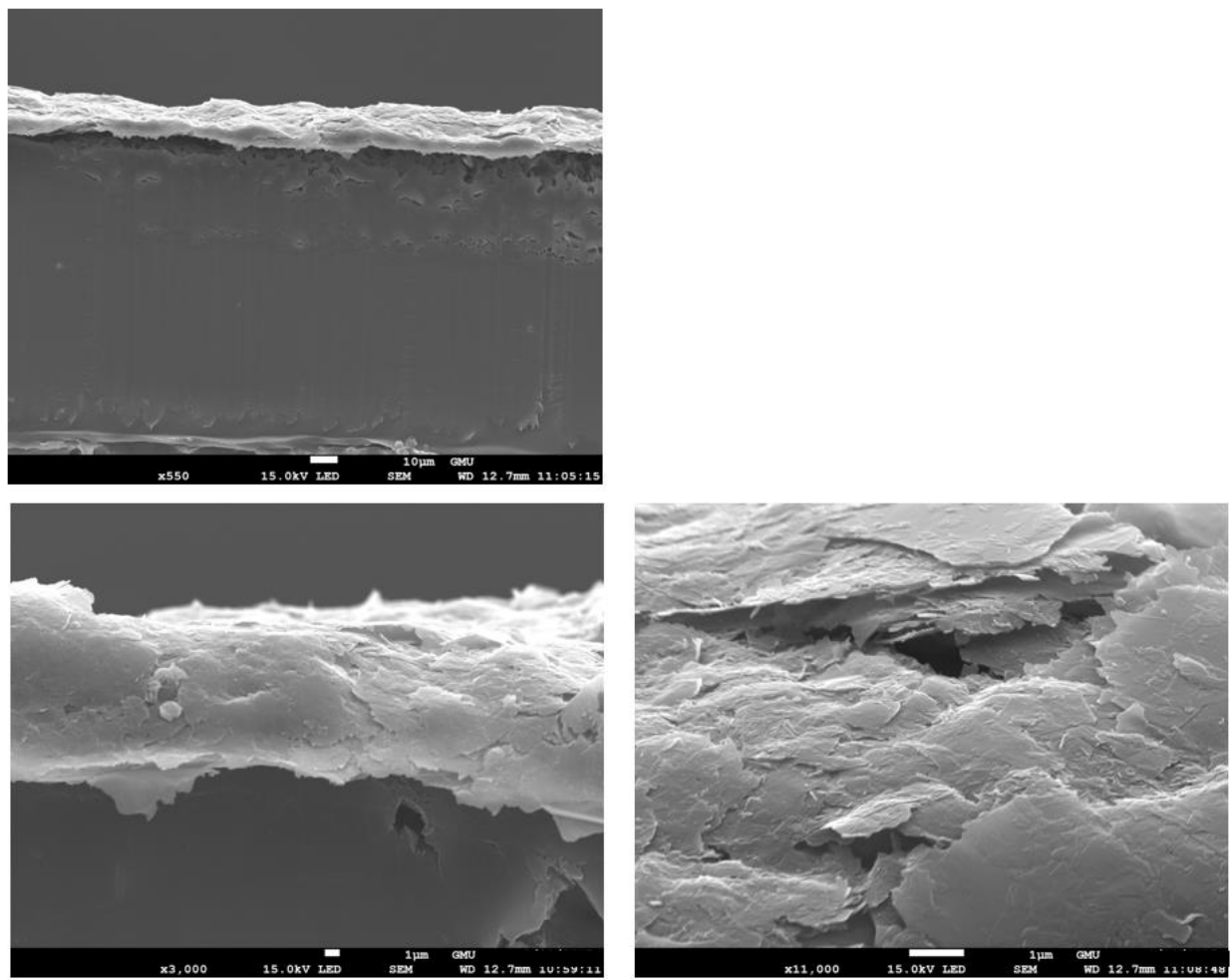

**Figure S1.** Cross-sectional SEM images of the graphene flake (GF).

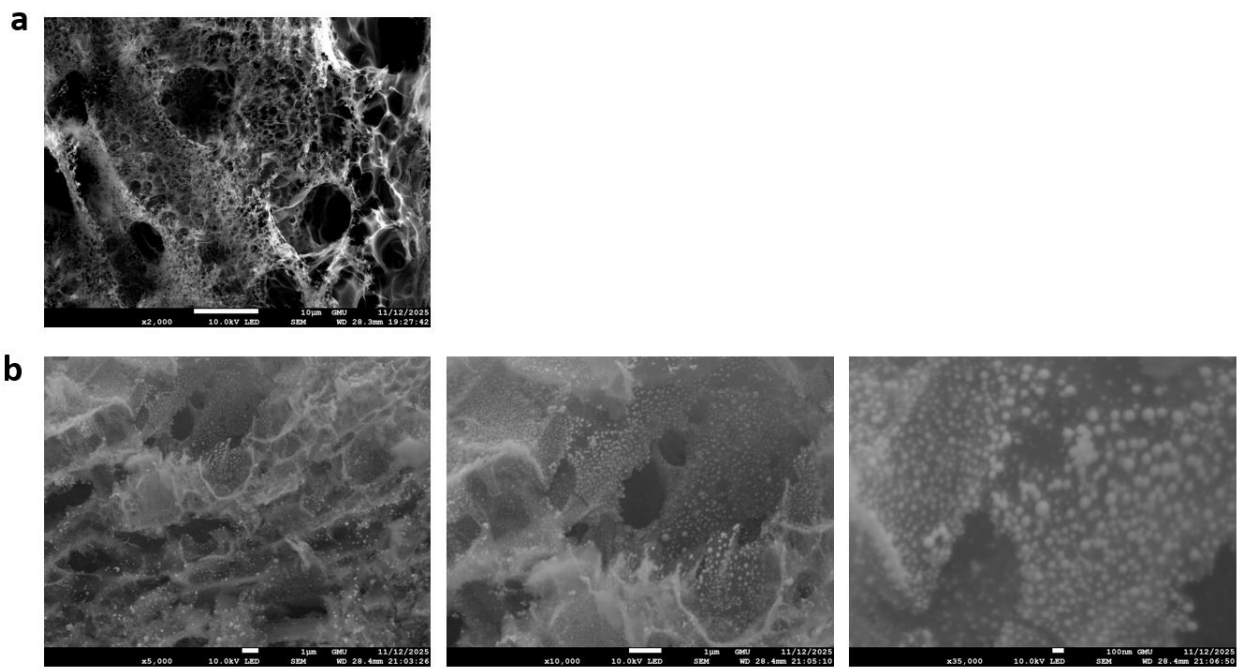

**Figure S2.** SEM images of the (a) LIG, (b) LIG-Au at different magnifications.

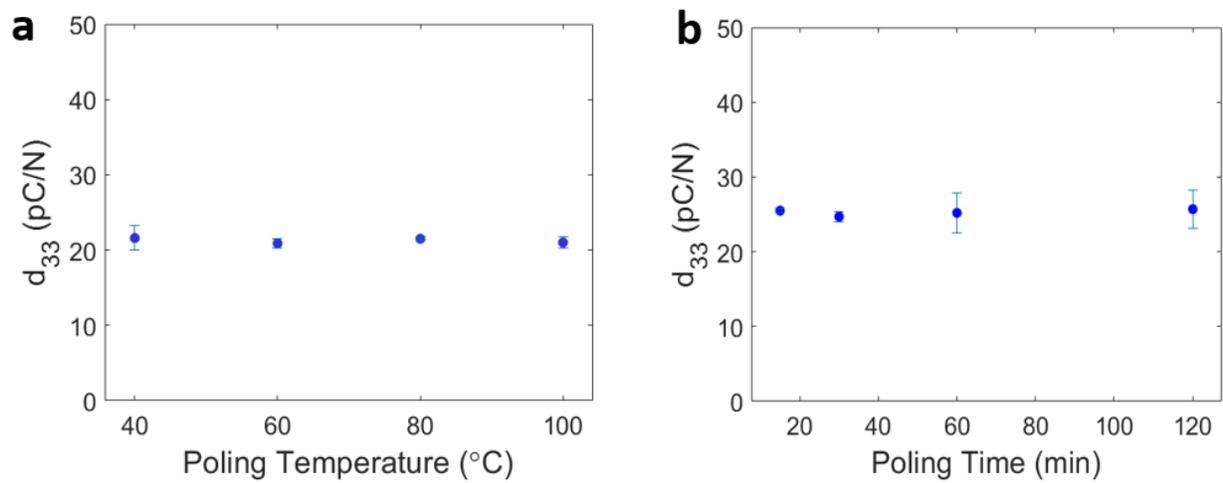

**Figure S3.** Piezoelectric charge coefficient ($d_{33}$) as a function of (a) poling temperature, and (b) poling duration.

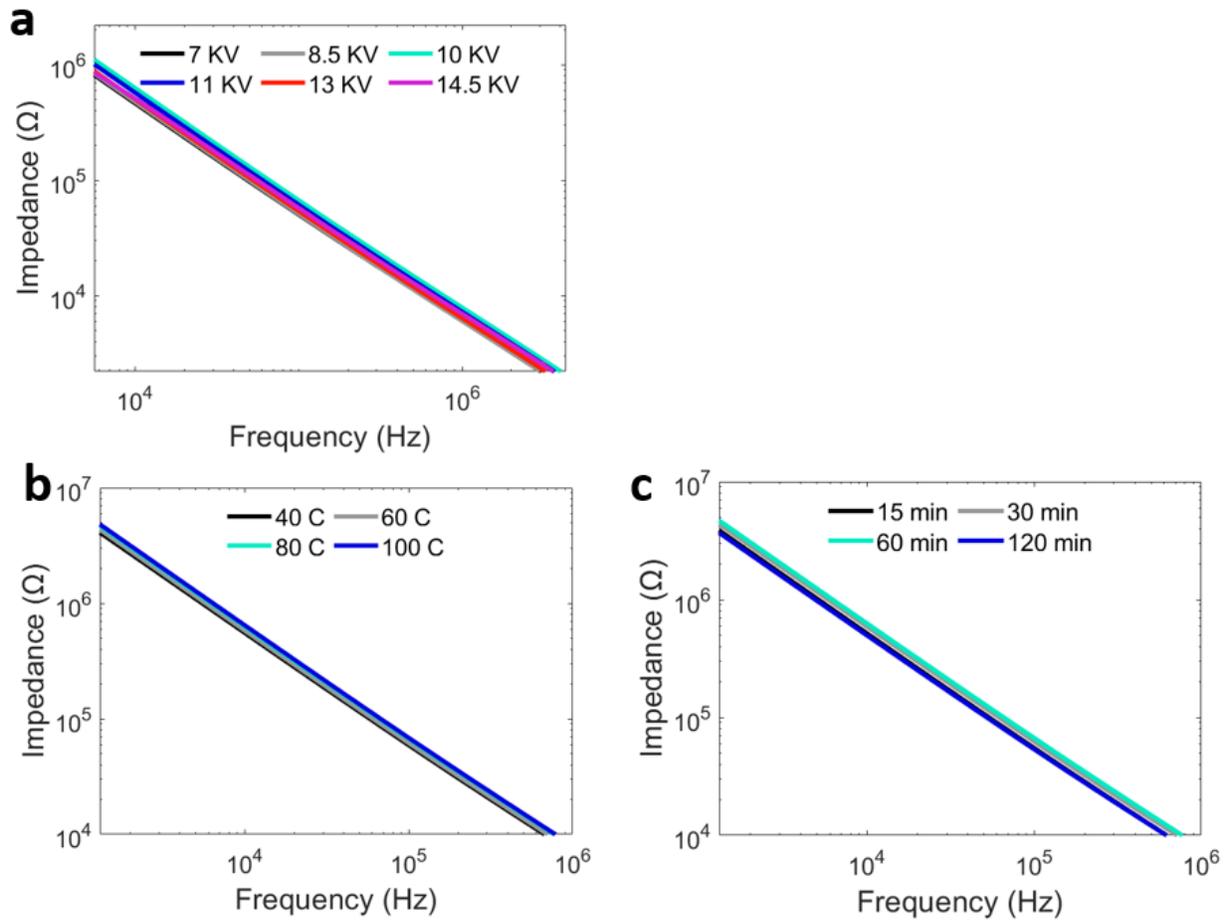

**Figure S4.** Impedance characteristics of PVDF-TrFE thin films under varying poling conditions. (a) Impedance as a function of applied electric field strength, measured at constant temperature and poling time. (b) Temperature-dependent impedance behavior, recorded at fixed electric field strength and poling duration. (c) Time evolution of impedance during the poling process, measured at constant temperature and electric field strength.

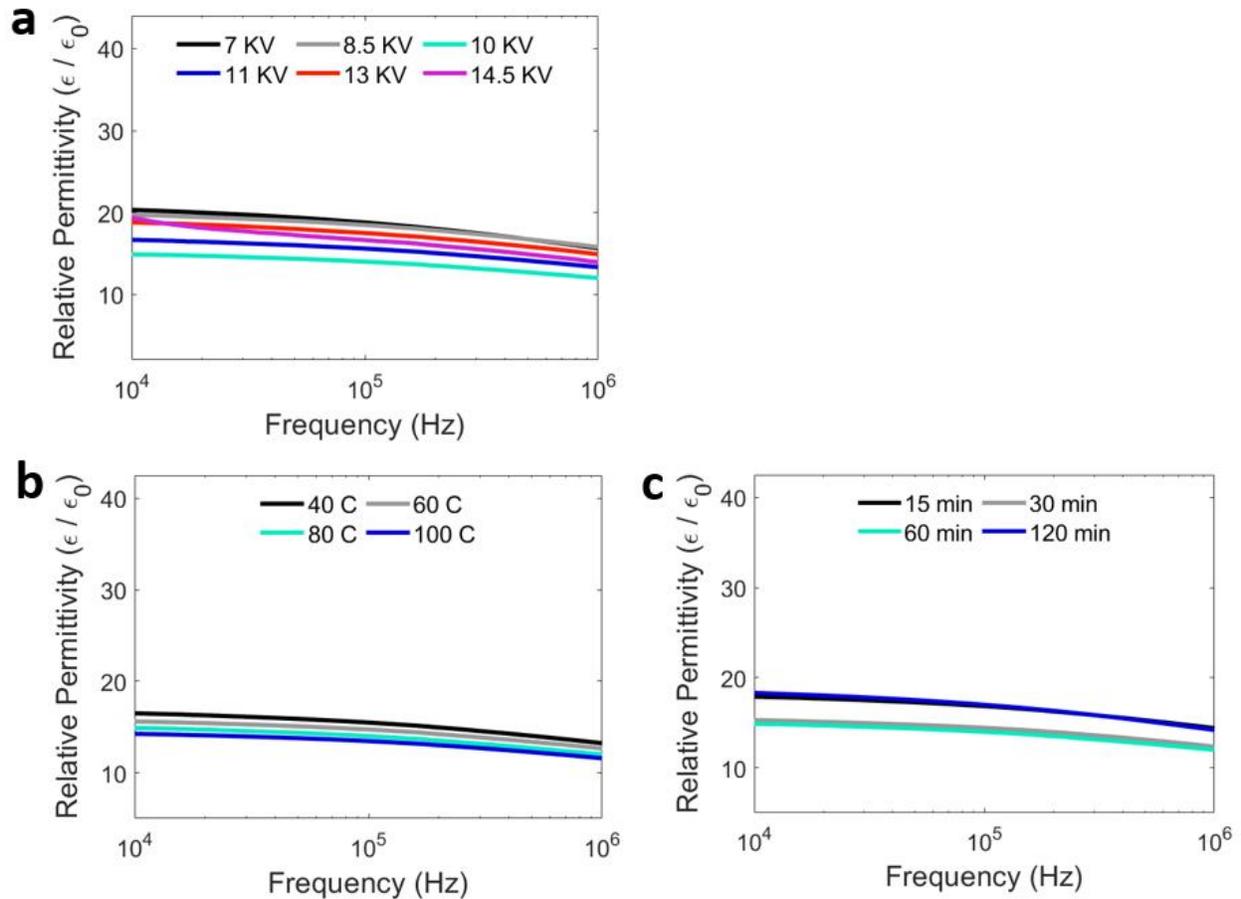

**Figure S5.** Dielectric properties of PVDF-TrFE thin films under various poling conditions. (a) Relative permittivity as a function of applied electric field strength, measured at constant temperature and poling time. (b) Temperature dependence of relative permittivity, recorded at fixed electric field strength and poling duration. (c) Time evolution of relative permittivity during the poling process, measured at constant temperature and electric field strength.

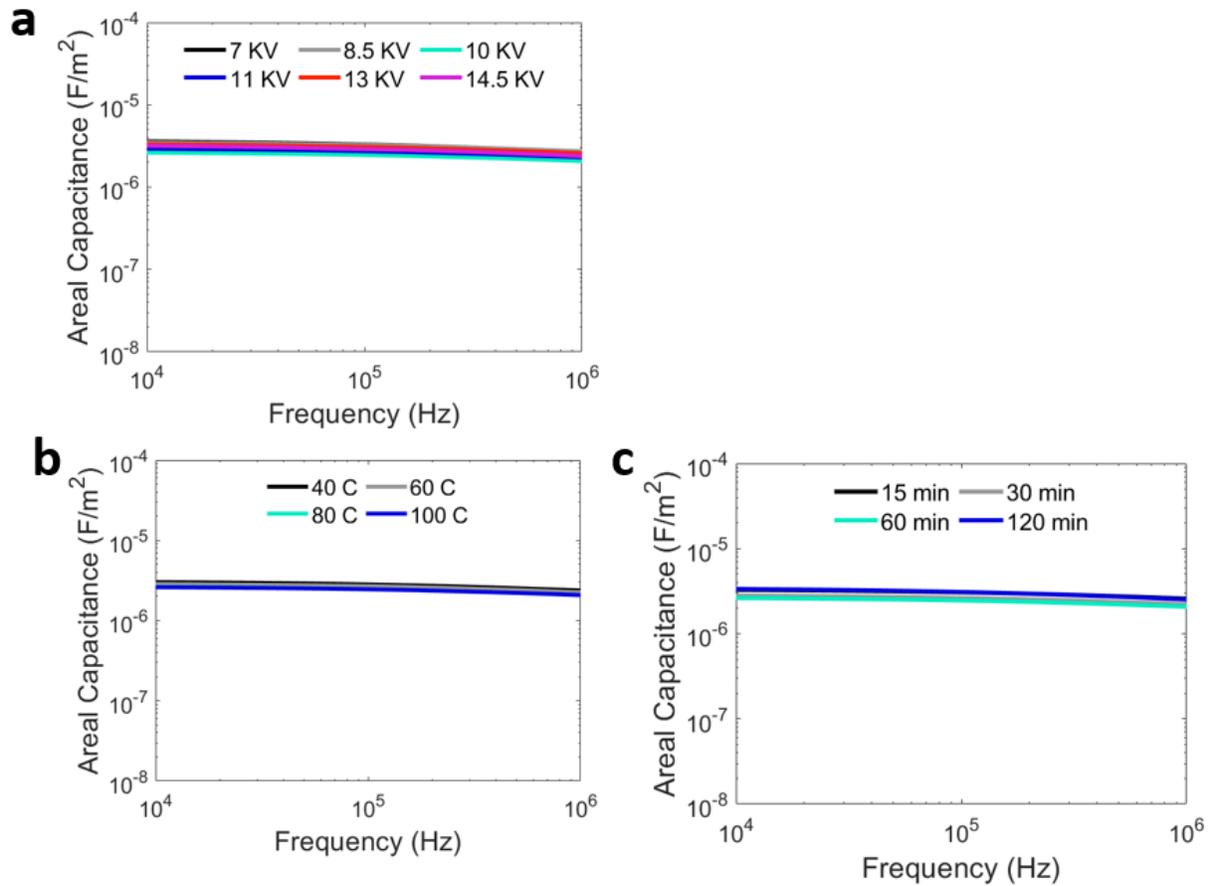

**Figure S6**. Capacitance behavior of PVDF-TrFE thin films under various poling conditions. (a) Areal capacitance as a function of applied electric field strength, measured at constant temperature and poling time. (b) Temperature dependence of capacitance, recorded at fixed electric field strength and poling duration. (c) Time evolution of capacitance during the poling process, measured at constant temperature and electric field strength.

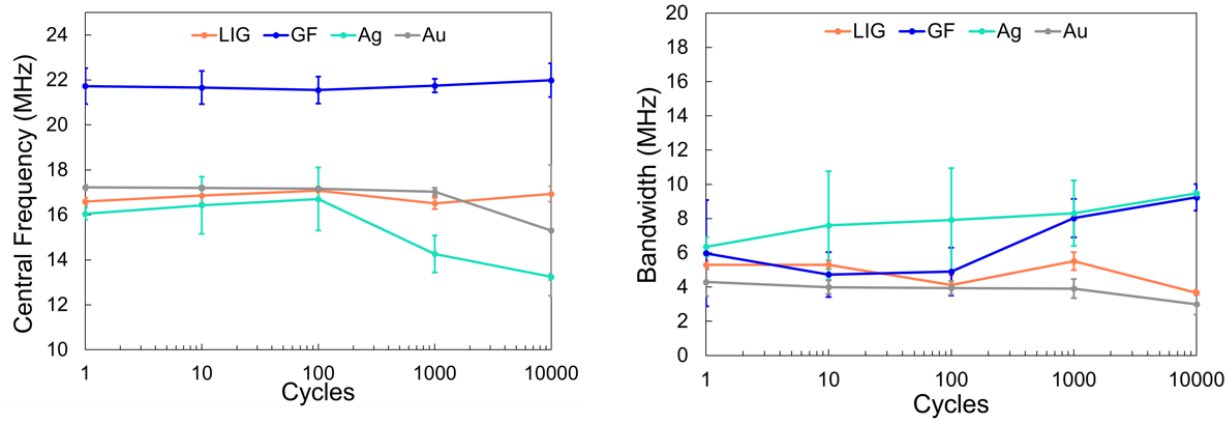

**Figure S7.** Central frequency and bandwidth of LIG-, GF-, Ag-, and Au-based transducers during cyclic bending up to 10,000 cycles.

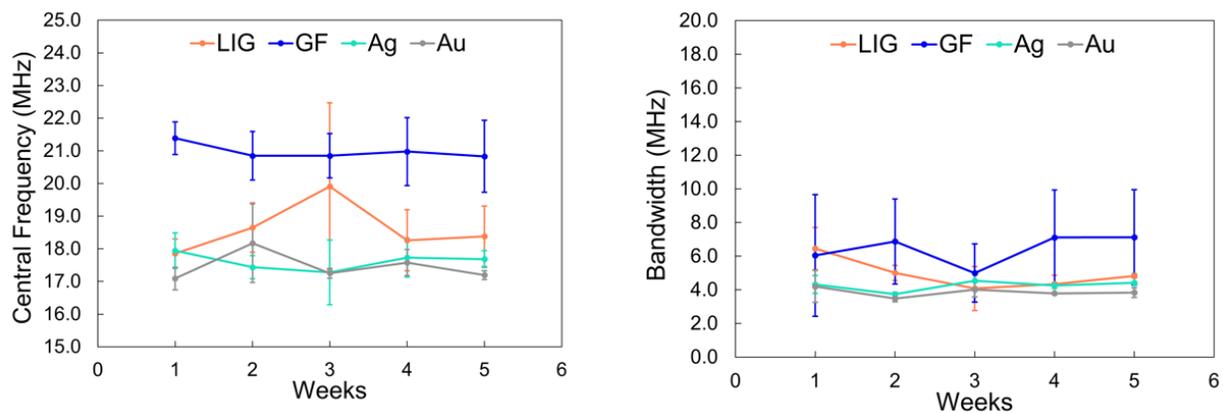

**Figure S8.** Central frequency and bandwidth of LIG-, GF-, Ag-, and Au-based transducers over

8 weeks aging test.

## 2. Supplementary Movies

**Movie S1.** The cyclic bending under 22.6 mm bending diameter up to 10,000 times.